# Thermodynamic properties of Aharanov-Bohm(AB) and magnetic fields with screened Kratzer potential

Akpan N.Ikot[1], Collins O. Edet[1], Precious O.Amadi[1], Uduakobong S.Okorie[1&2],

G.J.Rampho[3] and H.B.Abdullah[4]

[1] Theoretical Physics Group, Department of Physics, University of Port Harcourt, Choba, Nigeria

[2]Department of Physics, Akwa Ibom State University, Ikot Akpaden, P.M.B. 1167,Uyo. Nigeria.

[3]Department of Physics, University of South Africa-South Africa.

[4]Department of Physics, College of Education, Salahaddin University, 44001, Erbil, Krg-Iraq

## Abstract

In this study, the Schrodinger equation (SE) with screened Kratzer potential (SKP) in the presence of external magnetic and AB-flux fields is investigated using the factorization method. The eigenvalue and eigenfunction for the system are obtained in closed form. It is found that the present of the magnetic field partially removes the degeneracy when the screening parameter of the potential was small $(\alpha = 0.005)$ but the addition of the AB field removed the degeneracy faster and better. The magnetization and magnetic susceptibility of the system are evaluated at zero and finite temperatures and other thermodynamic properties of the system are discussed. More so, the presence of the AB-flux field makes the system to exhibit a both a paramagnetic and diamagnetic behavior. A straight forward extension of these results to three dimension shows that the present result is consistent with those obtained in literature.

1. Introduction

In non-relativistic (NR) quantum mechanics(QM), the solutions of the wave equation (i.e. "Schrödinger equation(SE)") [1-4] for particles in motion influenced by a central potential field has great applicability. This is because the eigenvalues and eigenfunction contains altogether the crucial information regarding the quantum system [5, 6]. Many works concerning solutions of the Schrödinger equation with diverse quantum mechanical potentials have been carried out broadly in the literature [7-10]. One of the potentials is the "screened Kratzer" potential(SKP) [11] which was recently proposed by Ikot et al. [11]. The expression for the SKP is stated as follows [11]:

$$V(r) = \left(-\frac{A}{r} + \frac{C}{r^2}\right)e^{-\alpha r}, \tag{1}$$

the letter $A \equiv 2D_e r_e$ and $C \equiv D_e r_e^2$, $\alpha$ represents the "screening parameter", $D_e$ represents the "dissociation energy" and $r_e$ represents the "equilibrium bond length". This SKP is a general case of many potentials, it reduces to the "Kratzer potential" [12, 13] when $\alpha$ approaches zero, "Yukawa" (i.e. "screened Coulomb") potential [14,15] when $C = 0$, and the "Coulomb" potential [8, 16] when $C = 0$ and $\alpha \to 0$. More so, if $A = 0$, the potential shrinks to the inversely quadratic Yukawa potential [17, 18].

Ikot et al [11] in a maiden consideration carried out a "non-relativistic(NR)" treatment [4] with this potential. The "eigenvalue" and normalized wave function for this system potential was explicitly obtained for carefully chosen diatomic molecules. Thermal properties of this system was obatined. Copious reports considering the combined effect of magnetic and "Aharanov-Bohm flux" fields have been carried out by several researchers.

For example, de Souza et al [19] examined the impact of impurity on the quantum dynamics of huge excitations in disclinated "graphene" bearing an external magnetic field. The disclination is brought in using an expression in the "Dirac equation" that translates the appearance of a phase associated with the transport of the "spinor" around the apex of the cone. The Dirac equation for this problem was exactly solved and the energy equations were obtained [19]. The influence of "screw dislocation (SD)" on the energy spectra and the eigenfunctions of an electron trapped in a two-dimensional pseudo-harmonic "quantum dot" [20] under the influence of an external magnetic field($\vec{B}$-field) inside a dot and Aharonov–Bohm field inside a pseudodot was investigated by Filgueiras et al [20]. The NR equation was solved analytically, the energy equation and eigenfunctions were calculated [19].

Ferkous et al. [21] examined the "bound state" energies for a fermionic massive particle with $\frac{1}{2}$ spin exposed to the "modified Poschl-Teller(PT) potential" and an AB field in the NR limit. The delta-function singularity problem which depicts the relationship between spin and magnetic flux was solved[21]. Spectral properties of an electron in two-dimensional (2D) "Gaussian quantum dot" (GQD) [22] under the collective effect of $\vec{B}$-field, electric field($\vec{E}$-field), and AB-flux field was studied by Aalu theoretically [23]. The problem was solved via the Nikiforov–Uvarov (NU) method [24] within the effective-mass approximation and the results were found and compared with that of the parabolic potential (PP) model [25].

Dantas et al. [26] explored the influence of the rotation of an electron and hole with SD restrained in a quantum ring(QR) potential. The Tan–Inkson(TI) potential [27, 28] was applied to model the confinement of the particle in a $2D$ QR. The authors assumed that the QR was situated in a $\vec{B}$-field and AB –flux filed in the center of the system, and the frame rotates about the *z*-axis. The SE was solved and the eigenfunctions and energy eigenvalues. The impact of the dislocation and the rotation the persistent current and magnetization was examined.

For the first time, Bahar and Soylu[29] obtained the eigenenergies of two-electron "Hulthen" [30, 31] QDs embedded in "Debye" and quantum plasmas modelled by the more general exponential cosine screened Coulomb (MGECSC) potential[32] under the combined influence of $\vec{E}$ and $\vec{B}$ fields. The modified SE was solved numerically via the asymptotic iteration method (AIM) [33]. It was posited that plasma conditions form a considerable QM effects for QDs and other atomic systems and that plasmas are pertinent experimental arguments.

A crucial and fascinating problem in physics is to study the thermal and magnetic properties of quantum systems. Some researchers have carried investigations in this direction. For example, Jahan et al. [34] scrutinized the effect geometry on the electronic, thermal, magnetic, and transport properties of a Gallium Arsenide(GaAs) QD using the power-exponential potential model [35] with a steepness parameter. The thermal quantities were calculated using the canonical ensemble approach(CEA) at low temperature. Khordad [36] evaluated some thermal quantities of an asymmetric GaAs QD with the effect of temperature and $\vec{B}$-field. The energy eigenvalue and eigenfunction were obtained using a transformation to simplify the Hamiltonian of the system.

Again, Jahan et al. [37] studied the problem of an exciton trapped in a Gaussian QD of GaAs in both $2D$ and $3D$ in the presence of a $\vec{B}$-field using the Ritz variational method[38], the $1/N$ expansion method [39], and the shifted $1/N$ expansion method[40, 41]. The ground state energy and the binding energy of the exciton were obtained and compared with results in literature. Shaer et al. [42] calculated the magnetization and susceptibility of two electrons interrelating and confined in a QD in the presence of a $\vec{B}$-field. The problem was solved via the exact diagonalization method [43, 44]. In another consideration, Khordad and his collaborators in 2019 investigated the statistical properties of a GaAs QD with an effective-PP [45]. The selected statistical quantities of the QD evaluated via CEA in the presence of a $\vec{B}$-field and its interaction with the electron spin. The relationship that exists between the external fields and the various thermodynamic functions were duly analyzed [45].

In this report, the purpose is in four-fold. First, the work of Ikot et al [11] is extended and the Schrödinger wave equation (SWE) is solved with the SKP model in the presence of external magnetic and AB-flux fields. The functional analysis approach (FAA) is used to obtain detailed solutions of the 2D SWE with the SKP model in the presence of magnetic and AB-flux fields. The resulting energy equation is used to obtain the partition function, which will be used to obtain other statistical quantities, such as mean energy entropy, free energy, specific heat capacity, magnetization and magnetic susceptibility. The effect of the fields on these properties will be analyzed. Furthermore, magnetisation and magnetic susceptibility at zero temperature is also considered.

The outline of the paper is as follows: Section 2 provides the solutions of the 2D SE with the SKP and vector potential $\vec{A}$ under the influence of external magnetic and AB-flux fields. In Section 3, the computations of numerical energy spectrum under external fields are considered and the comparison with previous results is given when fields become zero. In Section 4, magnetization and magnetic susceptibility at zero temperature is considered. The behaviour of thermodynamics properties in the presence of external fields is then studied in Section 5. Finally, the paper ends with concluding remarks in Section 6.

## 2. Schrödinger equation with SKP with AB flux field and an external $\vec{B}$-field

The Hamiltonian operator of a particle that is charged and subjected to move in the SKP under the combined impact of AB flux field and $\vec{B}$-field can be written in cylindrical coordinates. Thus, the SE is written as in Ref. [46-48] taking into consideration the SKP.

$$\left(i\hbar\vec{\nabla} - \frac{e}{c}\vec{A}\right)^2 \psi(r,\phi,z) = 2\mu\left[E_{nm} - \left(-\frac{A}{r} + \frac{C}{r^2}\right)e^{-\alpha r}\right]\psi(r,\phi,z), \tag{2}$$

where $E_{nm}$ represents the energy level, $\mu$ is the effective mass of the system, the vector potential which is denoted by "$\vec{A}$" can be written as a combination of two terms $\vec{A} = \vec{A}_1 + \vec{A}_2$ having the azimuthal components [49] and external $\vec{B}$-field with $\vec{\nabla}\times\vec{A}_1 = \vec{B}, \vec{\nabla}\times\vec{A}_2 = 0$, where $\vec{B}$ is the magnetic field. $\vec{A}_1 = \frac{\vec{B}e^{-\alpha r}}{1-e^{-\alpha r}}\hat{\phi}$ and $\vec{A}_2 = \frac{\phi_{AB}}{2\pi r}\hat{\phi}$ represents the additional magnetic flux $\phi_{AB}$ created by a solenoid with $\vec{\nabla}\cdot\vec{A}_2 = 0$. The vector potential in full is written in a simple form as [47]; $\vec{A} = \left(0, \frac{\vec{B}e^{-\alpha r}}{1-e^{-\alpha r}} + \frac{\phi_{AB}}{2\pi r}, 0\right)$.

$$\frac{1}{r^2} = \frac{\alpha^2}{\left(1-e^{-\alpha r}\right)^2} \tag{3}$$

We assume a wave function in cylindrical coordinates as $\psi(r,\phi) = \frac{1}{\sqrt{2\pi r}}e^{im\phi}\rho_{nm}(r)$, where $m$ denotes the magnetic quantum number. Inserting this wave function, the vector potential into Eq. (2) and using the approximation proposed by Greene and Aldrich [50] with some simple algebraic calculations, we arrive at the following radial 2$^{nd}$ order DE:

$$\rho''_{nm}(r) + \left[ \frac{2\mu E_{nm}}{\hbar^2} + \frac{2\mu A\alpha e^{-\alpha r}}{\hbar^2(1-e^{-\alpha r})} - \frac{2\mu C\alpha^2 e^{-\alpha r}}{\hbar^2(1-e^{-\alpha r})^2} + \frac{2m\tau}{\hbar}\left(\frac{\alpha \vec{B} e^{-\alpha r}}{(1-e^{-\alpha r})^2}\right) - \frac{\tau^2 \vec{B}^2 e^{-2\alpha r}}{\hbar^2(1-e^{-\alpha r})^2} \right] \rho_{nm}(r)$$

$$-\left( \frac{\tau^2 \alpha \vec{B} \Phi_{AB} e^{-\alpha r}}{\hbar^2(1-e^{-\alpha r})^2 \pi} - \frac{\left[(m+\xi)^2 - \frac{1}{4}\right]\alpha^2}{(1-e^{-\alpha r})^2} \right)\rho_{nm}(r) = 0 \qquad (4a)$$

where

$$\tau = -\frac{e}{c}, \quad \phi_0 = \frac{hc}{e} \text{ and } \xi = \frac{\Phi_{AB}}{\phi_0}. \qquad (4b)$$

For Mathematical simplicity, let's introduce the following dimensionless notations;

$$-\varepsilon_{nm} = \frac{2\mu E_{nm}}{\hbar^2 \alpha^2}, \beta_1 = \frac{2\mu A}{\hbar^2 \alpha}, \beta_2 = \frac{2\mu C}{\hbar^2}$$

$$\delta_1 = \frac{2m\tau \vec{B}}{\hbar\alpha}, \delta_2 = \frac{\tau^2 \vec{B}^2}{\hbar^2 \alpha^2}, \delta_3 = \frac{\tau^2 \vec{B}\Phi_{AB}}{\hbar^2 \alpha \pi}$$

$$\gamma = (m+\xi)^2 - \frac{1}{4} \qquad (5)$$

Now using the factorization method [51, 52] with the following substitution $s = e^{-\alpha r}$ into Eq. (4), we can simply write Eq. (4) in the s-coordinate as follows;

$$\frac{d^2\rho_{nm}(s)}{ds^2} + \frac{1}{s}\frac{d\rho_{nm}(s)}{ds} + \frac{1}{s^2(1-s)^2}\left[\begin{array}{c}-(\varepsilon_{nm} - \beta_1 + \delta_2)s^2 + (2\varepsilon_{nm} - \beta_1 - \beta_2 + \delta_1 - \delta_3)s \\ -(\varepsilon_{nm} + \gamma)\end{array}\right]\rho_{nm}(s) = 0 \qquad (6)$$

If we consider the boundary conditions

$$s \Rightarrow \begin{cases} 0, & r \to \infty, \\ 1, & r \to 0, \end{cases} \quad \text{when} \quad \tag{7}$$

with $\rho_{nm}(s) \to 0$, we assume the following ansatz of the form;

$$\rho_{nm}(s) = s^{\lambda}(1-s)^{\sigma} f(s) \tag{8}$$

where

$$\lambda = \sqrt{\varepsilon_{nm} + \gamma} \tag{9}$$

$$\sigma = \frac{1}{2} + \sqrt{\frac{1}{4} + \beta_2 + \delta_2 + \delta_3 - \delta_1 + \gamma} \tag{10}$$

On substitution of Eq. (8) into Eq. (6), we obtain the following hypergeometric equation:

$$s(1-s)f''(s) + \left[(2\lambda + 1) - (2\lambda + 2\sigma + 1)s\right]f'(s) - \left[\left((\lambda + \sigma) - \sqrt{\varepsilon_{nm} - \beta_1 + \delta_2}\right)\left((\lambda + \sigma) + \sqrt{\varepsilon_{nm} - \beta_1 + \delta_2}\right)\right]f(s) = 0, \tag{11}$$

its solutions are nothing but the hypergeometric functions

$$f(s) = {}_2F_1(a,b;c;s) \tag{12}$$

where

$$\begin{aligned} a &= (\lambda + \sigma) - \sqrt{\varepsilon_{nm} - \beta_1 + \delta_2} \\ b &= (\lambda + \sigma) + \sqrt{\varepsilon_{nm} - \beta_1 + \delta_2} \\ c &= 2\lambda + 1 \end{aligned} \tag{13}$$

If the finiteness of the solutions is considered, the quantum condition is given by

$$(\lambda + \sigma) - \sqrt{\varepsilon_{nm} - \beta_1 + \delta_2} = -n, \; n = 0, 1, 2... \tag{14}$$

from which we obtain

$$\varepsilon_{nm} = -\gamma + \left[ \frac{\delta_2 - \beta_1 - \gamma - \left(n + \frac{1}{2} + \sqrt{\frac{1}{4} + \beta_2 + \delta_2 + \delta_3 - \delta_1 + \gamma}\right)}{2\left(n + \frac{1}{2} + \sqrt{\frac{1}{4} + \beta_2 + \delta_2 + \delta_3 - \delta_1 + \gamma}\right)} \right]^2 \tag{15}$$

Consequently, if one substitutes the value of the dimensionless parameters in Eq. (5) into Eq. (15), we obtain the solutions as follows:

$$E_{nm} = \frac{\hbar^2 \alpha^2}{2\mu}\left((m+\xi)^2 - \frac{1}{4}\right) - \frac{\hbar^2 \alpha^2}{2\mu} \left[ \frac{\frac{\tau^2 \vec{B}^2}{\hbar^2 \alpha^2} - \frac{2\mu A}{\hbar^2 \alpha} - (m+\xi)^2 + \frac{1}{4} - (n+v)^2}{2(n+v)} \right]^2 \tag{16a}$$

$$v = \frac{1}{2} + \sqrt{\frac{\tau^2 \vec{B}^2}{\hbar^2 \alpha^2} + \frac{2\mu C}{\hbar^2} - \frac{2m\tau \vec{B}}{\hbar \alpha} - \frac{\tau^2 \vec{B} \Phi_{AB}}{\hbar^2 \alpha \pi} + (m+\xi)^2} \tag{16b}$$

The $3D$ NR energy solutions are obtained by setting $m = \ell + \frac{1}{2}$, in Eq. (16) to obtain[11];

$$E_{n\ell} = \frac{\hbar^2 \alpha^2 \ell(\ell+1)}{2\mu} - \frac{\hbar^2 \alpha^2}{2\mu} \left[ \frac{\left(n + \frac{1}{2} + \sqrt{\frac{1}{4} + \ell(\ell+1)\frac{2\mu C}{\hbar^2}}\right)^2 + \ell(\ell+1) + \frac{2\mu A}{\hbar^2 \alpha}}{2\left(n + \frac{1}{2} + \sqrt{\frac{1}{4} + \ell(\ell+1)\frac{2\mu C}{\hbar^2}}\right)} \right]^2 \tag{17}$$

where $\ell$ is the rotational quantum number

The corresponding unnormalized wave function is obtain as

$$\rho_{nm}(s) = s^{\sqrt{\varepsilon_{nm}+\gamma}} (1-s)^{\frac{1}{2}+\sqrt{\frac{1}{4}+\beta_2+\delta_2+\delta_3-\delta_1+\gamma}} \,_2F_1\left(-n, n+2(\lambda+\sigma); 2\lambda+1; s\right) \tag{18}$$

### 3. Applications

In this section, the numerical and graphical results of this study is discussed. Table 1 shows the energy eigenvalue computed using Equation (16) for four different cases at $\alpha = 0.005$; when both $\vec{B} = \Phi_{AB} = 0$, there is the existence of degeneracy. By subjecting the system to only a magnetic field $(\vec{B} \neq 0, \Phi_{AB} = 0)$, the energy eigenvalues are increased and degeneracy is removed; however, there is still the existence of quasi-degeneracy. However, when only the AB field is applied $(\vec{B} = 0, \Phi_{AB} \neq 0)$, the degeneracy is completely removed in comparison to the effect of the magnetic field and the energy eigenvalues increases. Therefore, the total consequences of both fields are stronger than the single individual effects; thus, there is a major shift in the bound state energy for the system.

In Table 2, we compute the energy eigenvalue using Equation (16a) for four cases with an increased value of the screening parameter $(\alpha = 0.01)$. When both fields are absent (i.e., $B = 0, \Phi_{AB} = 0$), degeneracy is observed. Again, by subjecting the entire system to only the magnetic field, the energy values are reduced and degeneracy is quasi-removed, and the energy levels become more negative. When only the AB-flux is present, the degeneracy is removed swiftly and energy eigenvalue increases, which indicates that the system is strongly attractive. Moreover, the dual effect of the external fields is stronger than the individual effects; consequently, there is an ample shift in the bound state energy of the system.

Figure 1(a) shows the combined effects of the AB-flux and magnetic fields on the energy values of the screened Kratzer potential. The influence of the AB-flux field on the quantum system is great. In the absence of the AB-flux field, the energy is higher but the energy increases as the magnetic field varies. The relationship in this case shows an exponential increase in the energy eigenvalue. Figure 1(b) shows that the energy eigenvalues increase monotonically as the magnetic field is increased. Figures 1(c) and 1(d) show that the energy eigenvalue increases as $\Phi_{AB}$ is increased. However, there is a quasi-linear trend in the absence of a magnetic field.

Figure 2 shows the variation of the partition function with the AB-flux field $\beta$ and the magnetic field. Figures 2(a) and 2(b) show that the partition function of the system increases with increasing $\vec{B}$ and $\Phi_{AB}$ fields. Although, in Figure 2(b), when $\beta = 0.01$, the partition function is almost invariant. In Figures 2(c) and 2(d), we show the variation of the partition function with varying $\beta$ with different values of AB-flux field and magnetic field respectively. In both cases, the relationship shows a linear trend in the variation of patterns.

Figure 3 shows the variation of the magnetisation with AB-flux field, magnetic field and $\beta$. In Figure 3(a), a continuous increase is shown in the magnetisation as the magnetic field varies up to the region $3.5 \leq \vec{B} \leq 4.8$, where a saturation was observed in the three cases considered. Beyond this point, the magnetisation drops. Figure 3(b) shows that the magnetisation increases as AB-flux field increases in an almost linear pattern. It is also noted that the magnetisation decreases for increasing $\beta$ for the cases examined. In Figure 3(c), the magnetisation increases as $\beta$ does the same for $\vec{B} = 002T$ and $\vec{B} = 003T$, and is also low, manifesting only in the region of $M_{nm}(\vec{B}, \Phi_{AB}, \beta) < 0$. However, for $\vec{B} = 001T$, it is higher and manifests in the $M_{nm}(\vec{B}, \Phi_{AB}, \beta) > 0$.

In Figure 3(d), $M_{nm}(\vec{B}, \Phi_{AB}, \beta)$ decreases monotonically with increasing $\beta$. Figure 4 illustrates the variation of magnetic susceptibility, $\chi_m(\vec{B}, \Phi_{AB}, \beta)$, with $\vec{B}, \Phi_{AB}$ and $\beta$. It is clearly shown in Figure 4(a) that $\chi_m(\vec{B}, \Phi_{AB}, \beta)$ increases with varying $\vec{B}$. It also shows that the system is purely paramagnetic in nature for this case.

In Figure 4(b), the magnetic susceptibility decreases as $\Phi_{AB}$ increases. In this case, it exhibits a diamagnetic behaviour. For Figure 4(c), susceptibility increases with increasing $\beta$ when $\vec{B} = 0.02T$ and exhibits the opposite behaviour when $\vec{B} = 0.03T$. The system exhibits a dual nature in this case because it is para-diamagnetic in nature. In the case, $\vec{B} = 0.01T$, and the susceptibility remained the same in the region of $\chi_m(\vec{B}, \Phi_{AB}, \beta) < 0$. Moreover, in Figure 4(d), $\chi_m(\vec{B}, \Phi_{AB}, \beta)$ decreases with increasing $\beta$ linearly.

Figure 5 shows the plots of the internal energy $U(\vec{B}, \Phi_{AB}, \beta)$ with the AB-flux field, magnetic field, and $\beta$. In Figure 5(a), we notice that in the three cases considered, in the region $3.3T \leq \vec{B} \leq 4.9T$, the internal energy drops abruptly and then gradually increases. Figure 5(b) shows that the internal energy decreases with increasing AB-flux field linearly. In Figures 5(c) and (d), the internal energy is plotted against $\beta$ with varying $\vec{B}$ and $\Phi_{AB}$, and the internal energy increases with increasing $\beta$ for both cases considered.

Figure 6 shows the variation of the specific heat with $\vec{B}, \Phi_{AB}$ and $\beta$. In Figure 6(a), the specific heat remained constant in the region of $0 \leq B \leq 0.05T$. Here we observe a sharp rise in the specific heat as the magnetic field increases and drops immediately. In Figure 6(b), the specific heat capacity increases as $\Phi_{AB}$ increased. Figures 6 (c) and (d) show that the specific heat increases with increasing $\beta$ in all cases considered. For $\vec{B} = 0.03T$ and $\Phi_{AB} = 3$, $C_v(\vec{B}, \Phi_{AB}, \beta)$, shows an almost uniform trend as it remained invariant throughout. The variation of the free energy, $F(\vec{B}, \Phi_{AB}, \beta)$ with $\vec{B}, \Phi_{AB}$ and $\beta$ is shown in Figure 7. In Figures 7(a) and 7(b), $F(\vec{B}, \Phi_{AB}, \beta)$ decreases as $\vec{B}$ and $\Phi_{AB}$ increases. In Figures 7(c) and (7d), $F(\vec{B}, \Phi_{AB}, \beta)$ increases with increasing $\beta$ as $\vec{B}$ and $\Phi_{AB}$ vary.

Figure 8 shows the variation of the entropy, $S(\vec{B}, \Phi_{AB}, \beta)$ with $\vec{B}, \Phi_{AB}$ and $\beta$. In Figure 8(a), the entropy remained steady in the region of $0 \leq B \leq 0.05T$, and shows a sharp rise in the entropy as the magnetic field increases and drops immediately. In Figure 8(b), the entropy increases as $\Phi_{AB}$ does the same. Figures 8(c) and 8(d) show that the entropy also increases as the inverse temperature $\beta$ is increased. However, there is a sharp and continuous decrease of the entropy in the case where $\Phi_{AB} = 8$, as shown in Figure 8(d).

In Figure 9, we show the behaviour of the magnetisation and magnetic susceptibility at zero temperature. In Figures 9(a) and 9(b), we show the variation of the magnetisation with varying magnetic and AB-flux fields, respectively. It is shown that in both cases, the magnetisation decreases as the $\vec{B}$ and $\Phi_{AB}$ fields are increased. In Figures 9(c) and 9(d), the variation of the magnetic susceptibility with varying AB and magnetic fields is displayed. In Figure 9(c), it is clearly seen that the susceptibility increases monotonically with increasing AB-flux fields. It also shows that the magnetic susceptibility gets saturated for higher values of the AB-flux field in the region of $35 \leq \Phi_{AB} \leq 50$. Figure 9(d) explicitly illustrates that the magnetic susceptibility decreases with an increasing magnetic field.

**Table1:** Numerical eigenvalues for the SKP model under the influence of AB flux field and external $\vec{B}$-field with various values of $m$. The fitting parameters used are as follows: $\hbar = A = \mu = \phi = e = c = 1, 2C = 1$ and $\alpha = 0.005$. All values are in natural units

| $m$ | $n$ | $\vec{B}=0, \Phi_{AB}=0$ | $\vec{B}=4, \Phi_{AB}=0$ | $\vec{B}=0, \Phi_{AB}=4$ | $\vec{B}=4, \Phi_{AB}=4$ |
|---|---|---|---|---|---|
| 0 | 0 | -0.224453125 | -0.0000101592 | -0.027740778 | 0.0001966517714 |
|   | 1 | -0.082421125 | -0.000041345 | -0.019583291 | 0.0001808333134 |
|   | 2 | -0.043305166 | -0.000097367 | -0.014852416 | 0.0001401207451 |
|   | 3 | -0.027225347 | -0.000178134 | -0.011878189 | 0.00007460689558 |
| 1 | 0 | -0.139473878 | -0.000028872 | -0.020427549 | 0.0002930090943 |
|   | 1 | -0.061617341 | -0.00008491 | -0.015428064 | 0.0002520130983 |
|   | 2 | -0.035300487 | -0.00016569 | -0.012289179 | 0.0001862168779 |
|   | 3 | -0.023352861 | -0.00027113 | -0.010201018 | 0.00009571267932 |
| -1 | 0 | -0.139473878 | 0.000008601 | -0.041415309 | 0.00009976076045 |
|   | 1 | -0.061617341 | 0.000002329 | -0.026533035 | 0.0001091844533 |
|   | 2 | -0.035300487 | -0.000028872 | -0.018830334 | 0.00009361967227 |
|   | 3 | -0.023352861 | -0.000084910 | -0.014347073 | 0.00005315983329 |

**Table 2:** Numerical eigenvalues for the SKP model under the influence of AB flux field and external $\vec{B}$-field with various values of $m$. The fitting parameters used are as follows: $\hbar = A = \mu = \phi = e = c = 1, 2C = 1$ and $\alpha = 0.01$. All values are in natural units

| $m$ | $n$ | $\vec{B}=0, \Phi_{AB}=0$ | $\vec{B}=4, \Phi_{AB}=0$ | $\vec{B}=0, \Phi_{AB}=4$ | $\vec{B}=4, \Phi_{AB}=4$ |
|---|---|---|---|---|---|
| 0 | 0 | -0.226701389 | -0.000040648 | -0.03209692 | 0.000786316 |
|   | 1 | -0.0848845 | -0.000165135 | -0.02340324 | 0.000721113 |
|   | 2 | -0.045873724 | -0.000388322 | -0.01841899 | 0.00055677 |
|   | 3 | -0.029889043 | -0.000709481 | -0.01534527 | 0.000294021 |
| 1 | 0 | -0.142507312 | -0.000115352 | -0.02491432 | 0.001168469 |
|   | 1 | -0.064404298 | -0.000338662 | -0.01942501 | 0.001001876 |
|   | 2 | -0.038052784 | -0.000659943 | -0.01603745 | 0.000736893 |
|   | 3 | -0.026143059 | -0.001078473 | -0.01384463 | 0.000374247 |
| -1 | 0 | -0.142507312 | 0.000034421 | -0.04556147 | 0.000399885 |
|   | 1 | -0.064404298 | 0.000009258 | -0.03011213 | 0.000436608 |
|   | 2 | -0.038052784 | -0.000115353 | -0.02217152 | 0.00037343 |
|   | 3 | -0.026143059 | -0.000338664 | -0.01760814 | 0.000211096 |

## 4. Thermal Properties of Hellmann Potential with Magnetic and AB fields

The partition function can be calculated if we carry out a direct summation over all possible energy levels at a given temperature $T$ [53-54], this is algebraically shown below as;

$$Z(\beta) = \sum_{n=0}^{\varpi} e^{-\beta E_{nm}}, \beta = \frac{1}{k_B T} \quad (19)$$

Here, $k_B$ is the Boltzmann constant and $E_{nm}$ is energy of the nth bound state.

We can rewrite eq. (16) to be of the form

$$E_{nm} = P_1 - \frac{\hbar^2 \alpha^2}{2\mu} \left( \frac{P_2 - (n+v)^2}{2(n+v)} \right)^2 \quad (20)$$

$$P_1 = \frac{\hbar^2 \alpha^2}{2\mu} \left( (m+\xi)^2 - \frac{1}{4} \right); \quad P_2 = \frac{\tau^2 \vec{B}^2}{\hbar^2 \alpha^2} - \frac{2\mu A}{\hbar^2 \alpha} - (m+\xi)^2 + \frac{1}{4} \quad (21)$$

We substitute eq. (20) into eq. (19) to have

$$Z(\beta) = \sum_{n=0}^{\eta} e^{-\beta \left[ P_1 - \frac{\hbar^2 \alpha^2}{2\mu} \left( \frac{P_2 - (n+v)^2}{2(n+v)} \right)^2 \right]} \quad (22)$$

where, $\eta = -v + \sqrt{P_1} \pm \sqrt{P_1 - P_2} \quad (23)$

is the maximum quantum number.

In the classical limit, the sum in Eq.(22) can be replaced by an integral, such that

$$Z(\beta) = \int_0^\eta e^{\beta \left( M(n+\eta)^2 + \frac{N}{(n+\eta)^2} + W \right)} dn \quad (24)$$

where

$$M = \frac{\hbar^2 \alpha^2 P_2^2}{8\mu}; N = \frac{\hbar^2 \alpha^2}{8\mu}; R = -\frac{\hbar^2 \alpha^2 P_2}{4\mu} - P_1. \quad (25)$$

$$Z(\beta) = \int_v^{\eta+v} e^{\beta \left( \frac{N}{\varphi^2} + M\varphi^2 + W \right)} d\varphi, \varphi = n + v \quad (26)$$

The integral is evaluated in the region $v \leq \varphi \leq \eta + v$

We therefore use Mathematica software to evaluate the integral in eq. (26), thus obtaining the partition function for the Hellmann potential model as;

$$Z(\beta) = \frac{-e^{W\beta - 2\sqrt{-M\beta}\sqrt{-N\beta}}\sqrt{\pi}\left(-Erf[\Lambda_1 - \Lambda_2] + e^{4\sqrt{-M\beta}\sqrt{-N\beta}}\left(Erf[\Lambda_1 + \Lambda_2] - Erf[\Pi_1 + \Lambda_2 + \Pi_2]\right) - \Xi\right)}{4\sqrt{-M\beta}} \quad (27a)$$

where we have used the following notations for mathematical simplicity

$$\Lambda_1 = \frac{\sqrt{-N\beta}}{v}, \Lambda_2 = \sqrt{-M\beta}v, \Pi_1 = \sqrt{-M\beta}\eta, \Pi_2 = \frac{\sqrt{-N\beta}}{\eta + v} \text{ and } \Xi = Erf[\Pi_1 + \Lambda_2 - \Pi_2] \quad (27b)$$

The error function can be defined as [55]

$$erf(z) = \frac{2}{\sqrt{\pi}} \int_0^z e^{t^2} dt \quad (28)$$

Thermodynamic functions such as; Magnetic Susceptibility, $\chi_m(\vec{B}, \Phi_{AB}, \beta)$ Helmholtz free energy, $F(\vec{B}, \Phi_{AB}, \beta)$, entropy, $S(\vec{B}, \Phi_{AB}, \beta)$, internal energy, $U(\vec{B}, \Phi_{AB}, \beta)$, and specific heat, $C_v(\vec{B}, \Phi_{AB}, \beta)$, functions can be obtained from the partition function(27) as follows;

**Magnetization at Finite Temperature**

The magnetization is given as[56];

$$M(\vec{B}, \Phi_{AB}, \beta) = \frac{1}{\beta}\left(\frac{1}{Z(\vec{B}, \Phi_{AB}, \beta)}\right)\left(\frac{\partial}{\partial \vec{B}} Z(\vec{B}, \Phi_{AB}, \beta)\right) \quad (29)$$

**Magnetic Susceptibility**;

The magnetic susceptibility of the system is calculated with [56]

$$\chi_m(\vec{B}, \Phi_{AB}, \beta) = \frac{\partial M(\vec{B}, \Phi_{AB}, \beta)}{\partial \vec{B}} \quad (30)$$

**Internal Energy**

The internal energy of the system is obtained as [57];

$$U(\vec{B}, \Phi_{AB}, \beta) = -\frac{\partial \left(\ln Z(\vec{B}, \Phi_{AB}, \beta)\right)}{\partial \beta} \quad (31)$$

**Specific Heat Capacity**

The specific heat capacity of is evaluated using the equation [57];

$$C_v(\vec{B}, \Phi_{AB}, \beta) = k_\beta \frac{\partial U(\vec{B}, \Phi_{AB}, \beta)}{\partial \beta} \quad (32)$$

**Free Energy**

The free energy of the system is given as [57]

$$F(\vec{B}, \Phi_{AB}, \beta) = -\frac{1}{\beta} \ln Z(\vec{B}, \Phi_{AB}, \beta) \tag{33}$$

**Entropy**

The entropy of the system is evaluated with the expression below [57];

$$S(\vec{B}, \Phi_{AB}, \beta) = -k_\beta \frac{\partial F(\vec{B}, \Phi_{AB}, \beta)}{\partial \beta} \tag{34}$$

## 5. Magnetization and Magnetic susceptibility at Zero Temperature.

In the present study, we are interested in analyzing the magnetization and magnetic susceptibility at zero temperature.

### 5.1 Magnetization

The magnetization of a system in a state $(n, m)$ are defined by[58];

$$M_{nm}(\vec{B}, \Phi_{AB}) = -\frac{\partial E_{nm}}{\partial \vec{B}} \tag{35}$$

### 5.2 Magnetic Susceptibility at zero temperature

The magnetic susceptibility at zero temperature is given as[58];

$$\chi_m = \frac{\partial M}{\partial \vec{B}} \tag{36}$$

6. **Conclusions**

In this paper, the effects of external and AB-flux fields on the energy spectra and thermodynamic properties with SKP were studied and analyzed. The factorization method was used to obtain the energy spectra and wave function for the system. The influence of the fields on the energy spectra of the system was analyzed. Furthermore, the magnetization and magnetic susceptibility of the system was considered at zero and finite temperatures. We evaluated the partition function and used it to evaluate other thermodynamic properties of the system. A comparison of the magnetic susceptibility of the system at zero and finite temperature shows similarity in the behaviour of the system. The presence of the AB-flux field makes the system exhibit a dual magnetic behaviour (paramagnetic and diamagnetic). These research findings could be applied in condensed matter physics, atomic physics, and chemical physics.

**Conflicts of interest**
The authors declare that there is no conflict of interest.

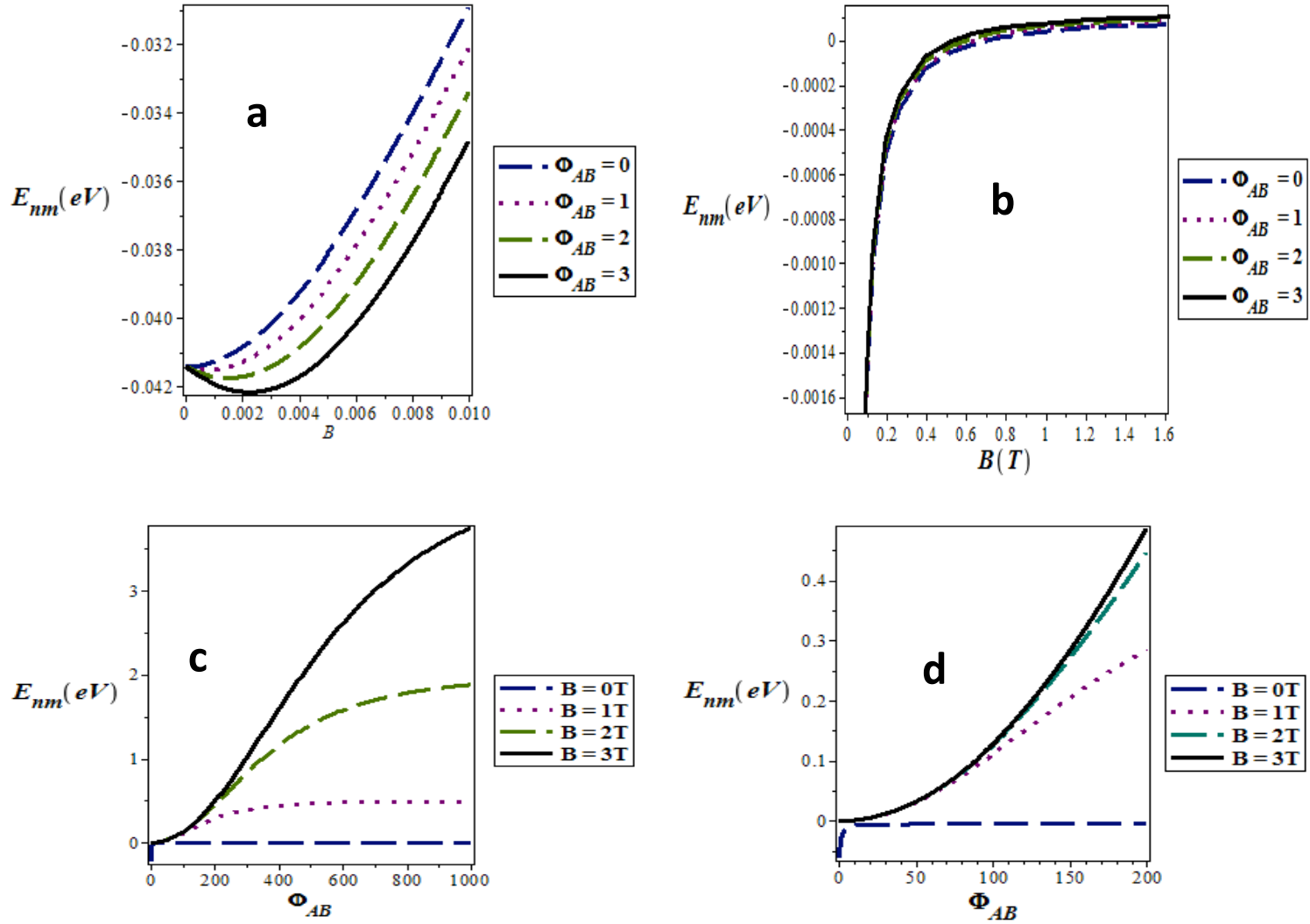

**Figure 1**: Variation of energy values for the SKP and under the influence of the $\vec{B}$-field and the AB flux field in natural units using the fitting parameters $\hbar = b = \mu = \phi = e = 2C = A = 1$ and $\alpha = 0.005$ **(a)** as a function of external $\vec{B}$-field with various $\Phi_{AB}$ and $m = n = 0$. **(b)** Same as (a) but with $m = n = 1$. **(c)** Variation of energy values for the SKP model and under the influence of the $\vec{B}$-field and the AB flux field in natural units using the fitting parameters $\hbar = b = \mu = \phi = e = c = 1$, $a = 2$ and $\alpha = 0.005$ as a function of AB flux field with various $\vec{B}$ and $m = n = 0$. **(d)** Same as (c) but with $m = n = 1$.

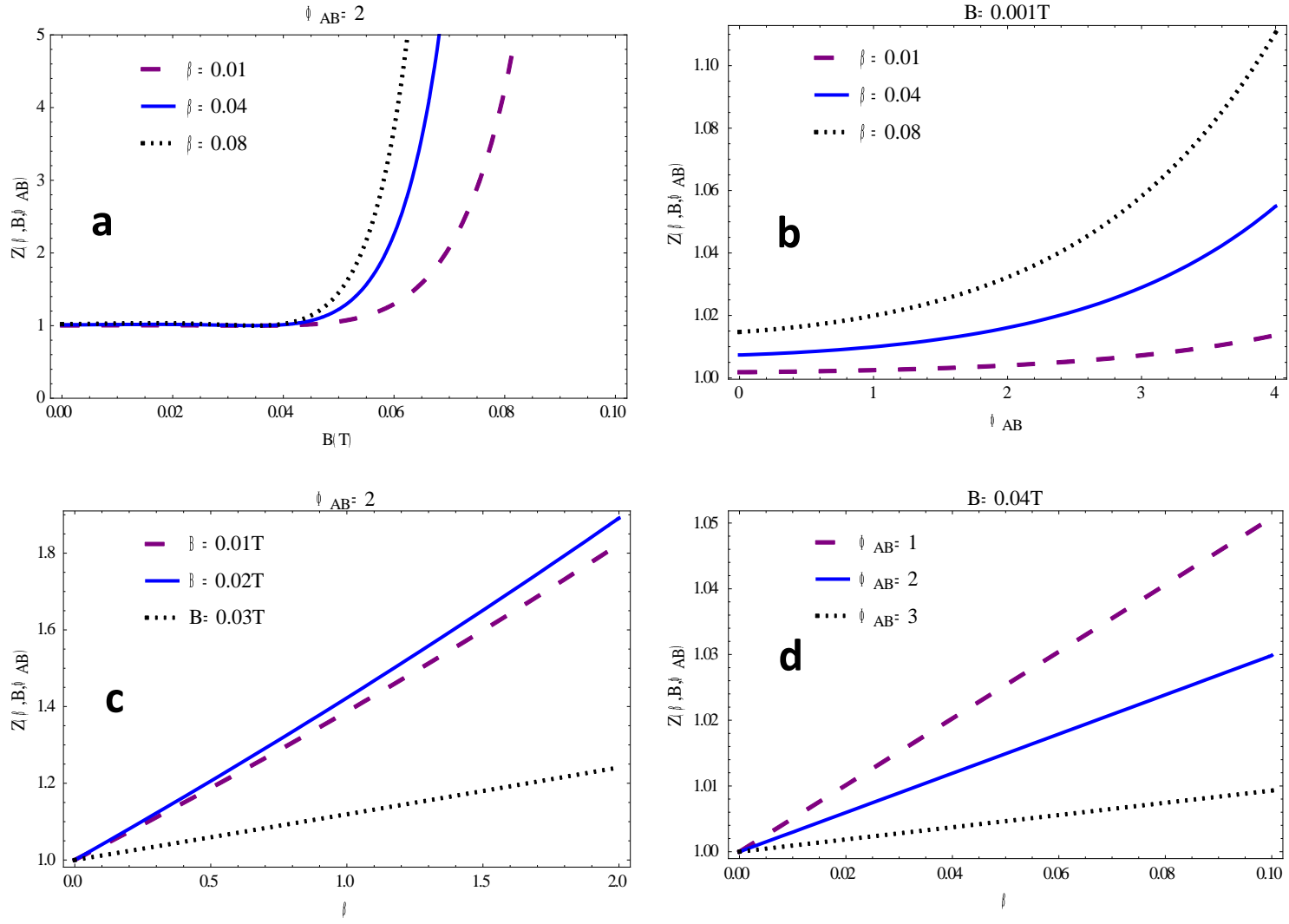

**Figure 2**; **(a)** Plot of Partition function against $\vec{B}$-field for different values of temperature. **(b)** Plot of Partition function against AB flux field for different values of temperature. **(c)** Plot of Partition function against $\beta$ for different values $\vec{B}$-field. **(d)** Plot of Partition function against $\beta$ for different values AB flux field, $\Phi_{AB}$.

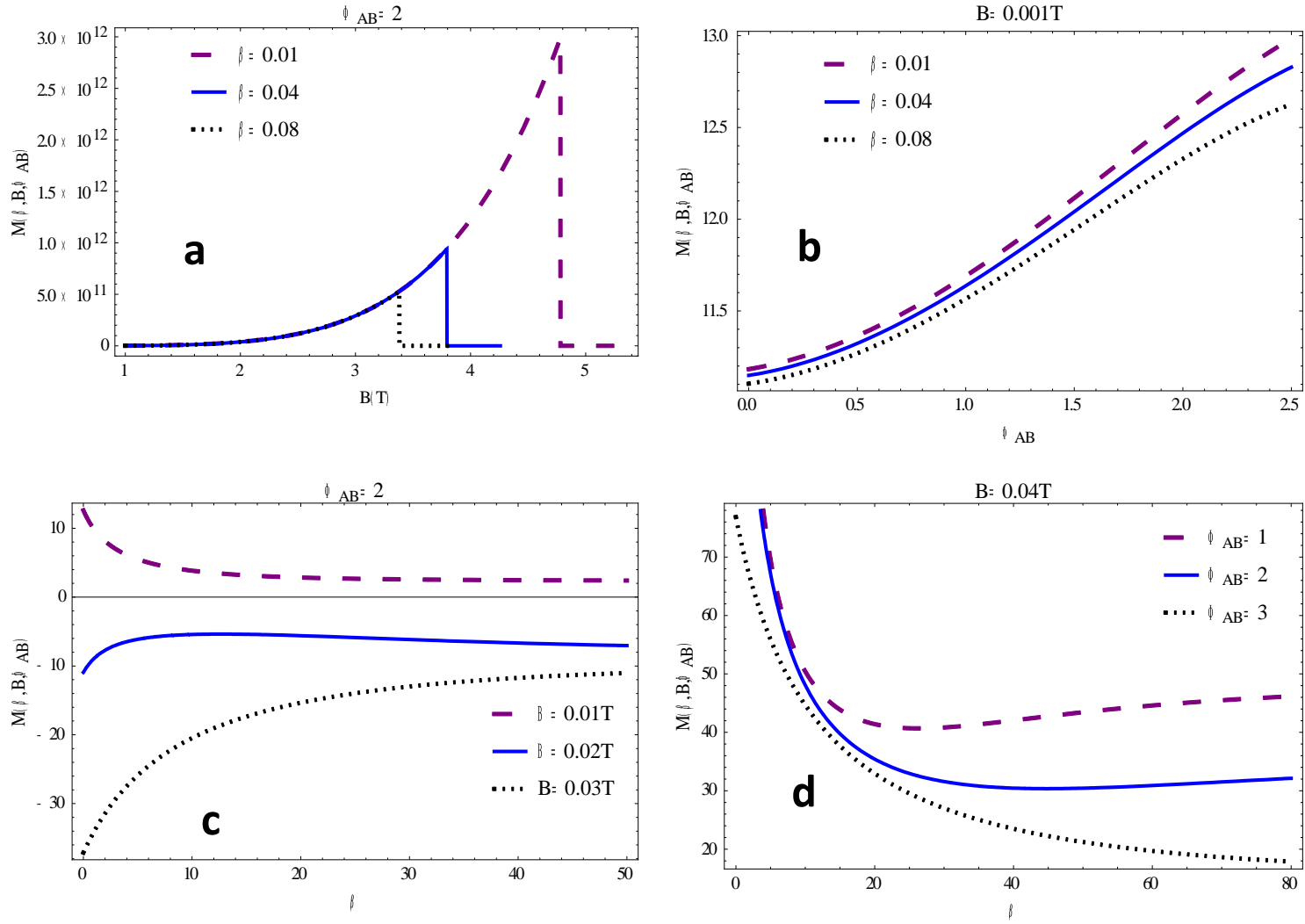

**Figure 3**; (a) Plot of Magnetization against $\vec{B}$-field at finite temperature. (b) Magnetization, $M\left(\vec{B}, \Phi_{AB}, \beta\right)$ against AB flux field, $\Phi_{AB}$ with different $\beta$. (c) Magnetization, $M\left(\vec{B}, \Phi_{AB}, \beta\right)$ against $\beta$ varying $\vec{B}$-field. (d) Magnetization, $M\left(\vec{B}, \Phi_{AB}, \beta\right)$ against $\beta$ varying $\vec{B}$-field.

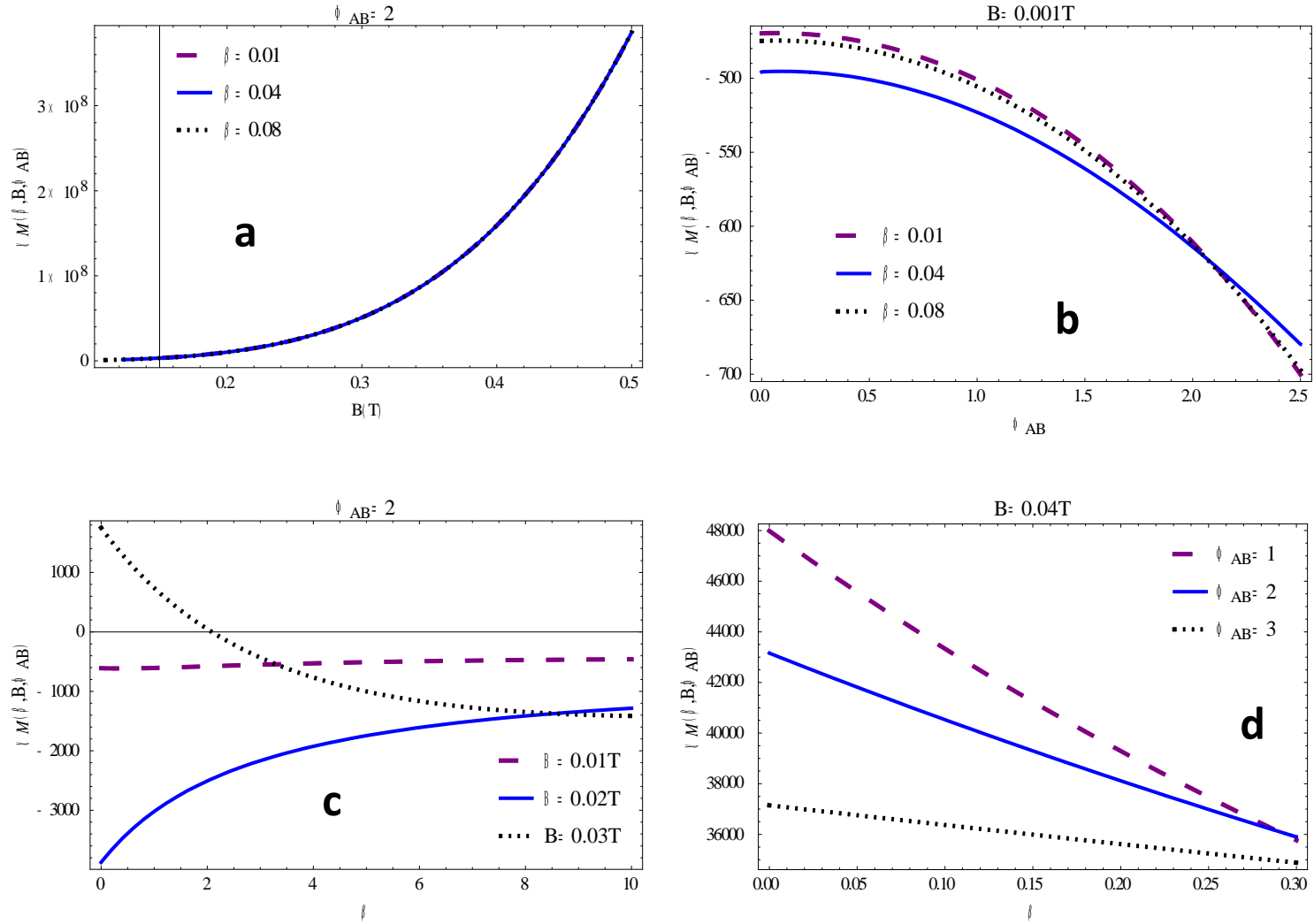

**Figure 4**; **(a)** Magnetic Susceptibility, $\chi_m\left(\vec{B},\Phi_{AB},\beta\right)$ against $\vec{B}$-field varying $\beta$. **(b)** Magnetic Susceptibility, $\chi_m\left(\vec{B},\Phi_{AB},\beta\right)$ against $\Phi_{AB}$ varying $\beta$. **(c)** Magnetic Susceptibility, $\chi_m\left(\vec{B},\Phi_{AB},\beta\right)$ against $\beta$ varying $\vec{B}$-field. **(d)** Magnetic Susceptibility, $\chi_m\left(\vec{B},\Phi_{AB},\beta\right)$ against $\beta$ varying $\Phi_{AB}$.

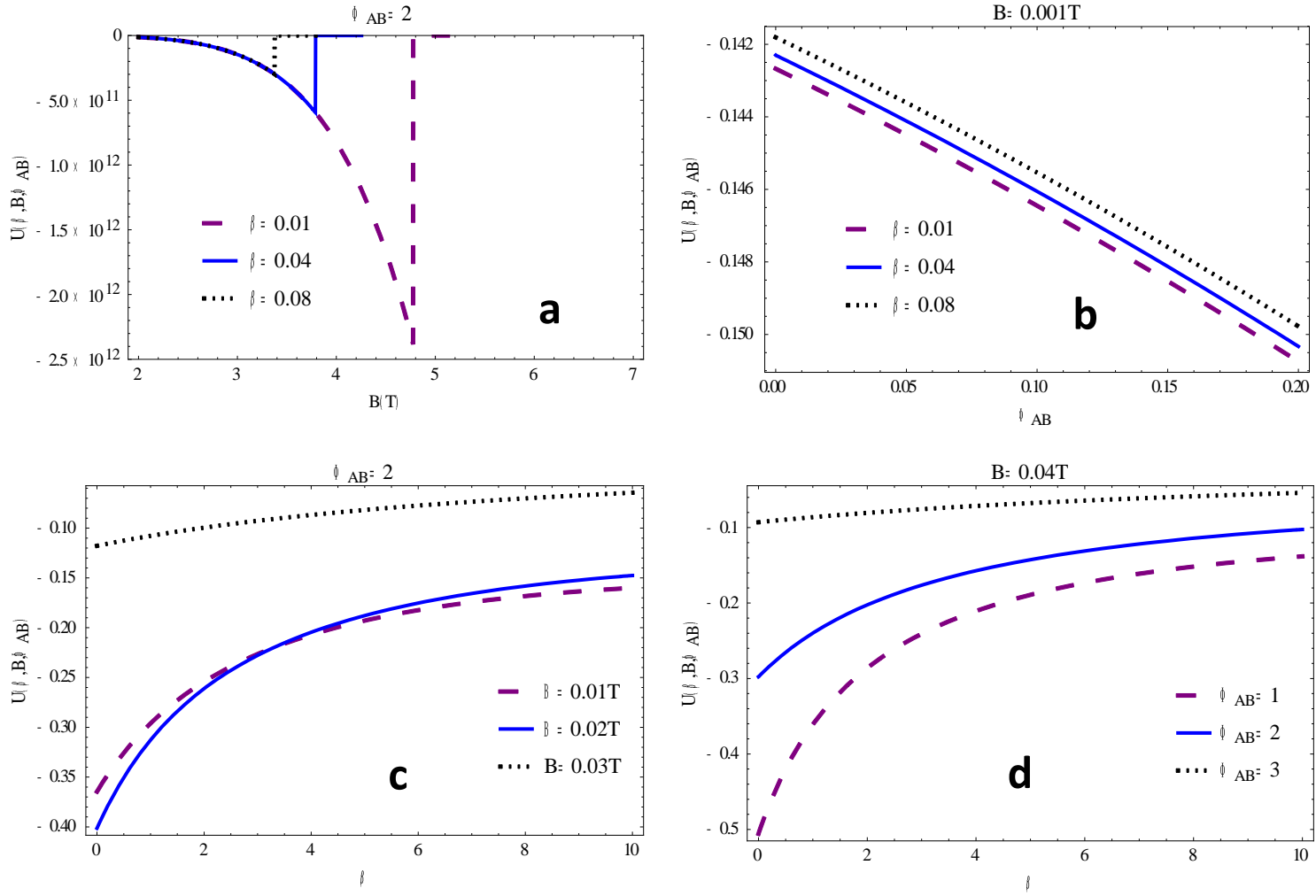

**Figure 5**; **(a)** Internal Energy, $U(\vec{B}, \Phi_{AB}, \beta)$ against $\vec{B}$-field varying $\beta$. **(b)** Internal Energy, $U(\vec{B}, \Phi_{AB}, \beta)$ against $\Phi_{AB}$ varying $\beta$. **(c)** Internal Energy, $U(\vec{B}, \Phi_{AB}, \beta)$ against $\beta$ varying $\vec{B}$-field. **(d)** Internal Energy, $U(\vec{B}, \Phi_{AB}, \beta)$ against $\beta$ varying $\Phi_{AB}$

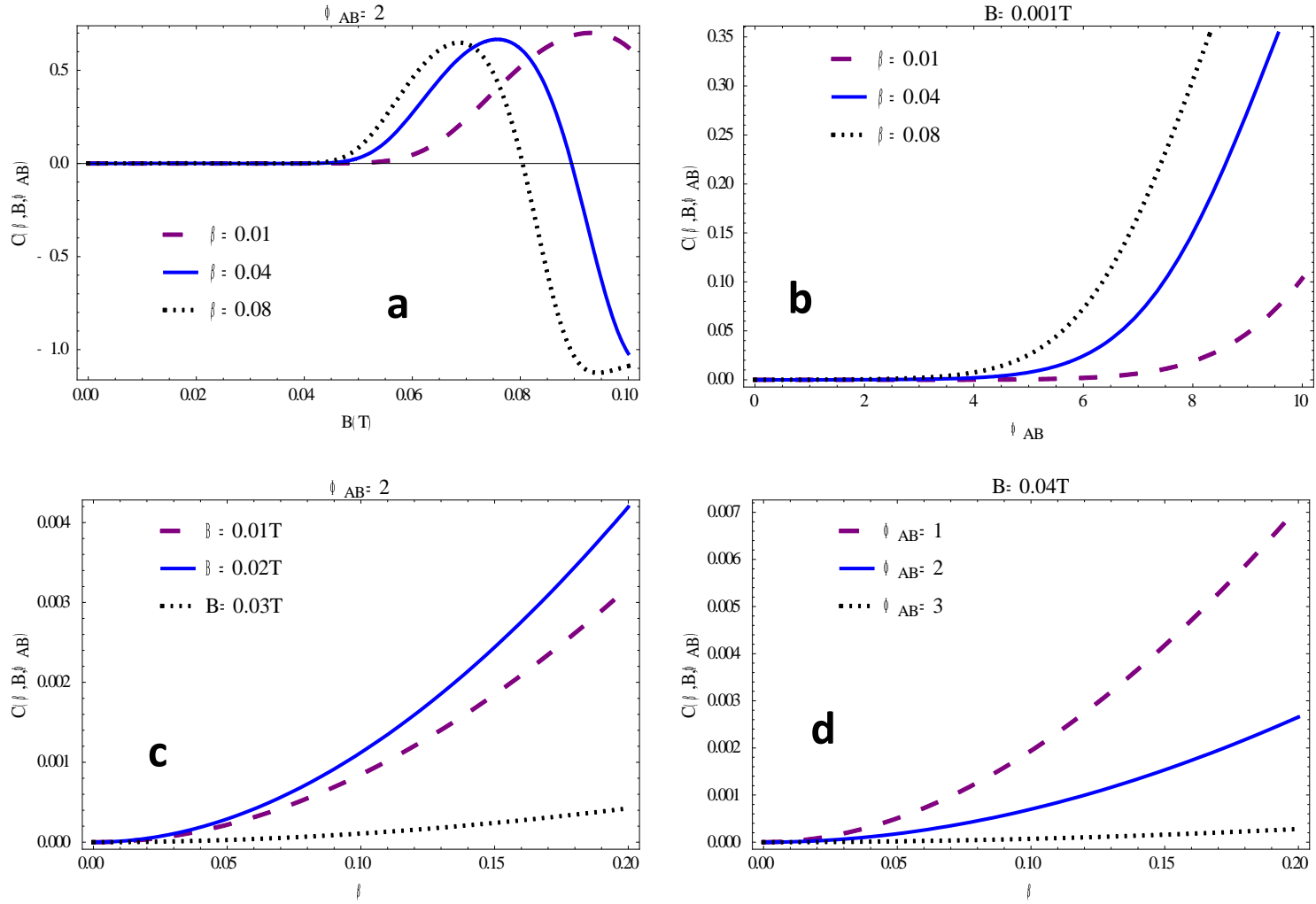

**Figure 6**; **(a)** Specific Heat Capacity, $C_v(\vec{B}, \Phi_{AB}, \beta)$ against $\vec{B}$-field varying $\beta$. **(b)** Specific heat capacity, $C_v(\vec{B}, \Phi_{AB}, \beta)$ against $\Phi_{AB}$ varying $\beta$. **(c)** Specific heat capacity, $C_v(\vec{B}, \Phi_{AB}, \beta)$ against $\beta$ varying $\vec{B}$-field. **(d)** Specific heat capacity, $C_v(\vec{B}, \Phi_{AB}, \beta)$ against $\beta$ varying $\Phi_{AB}$.

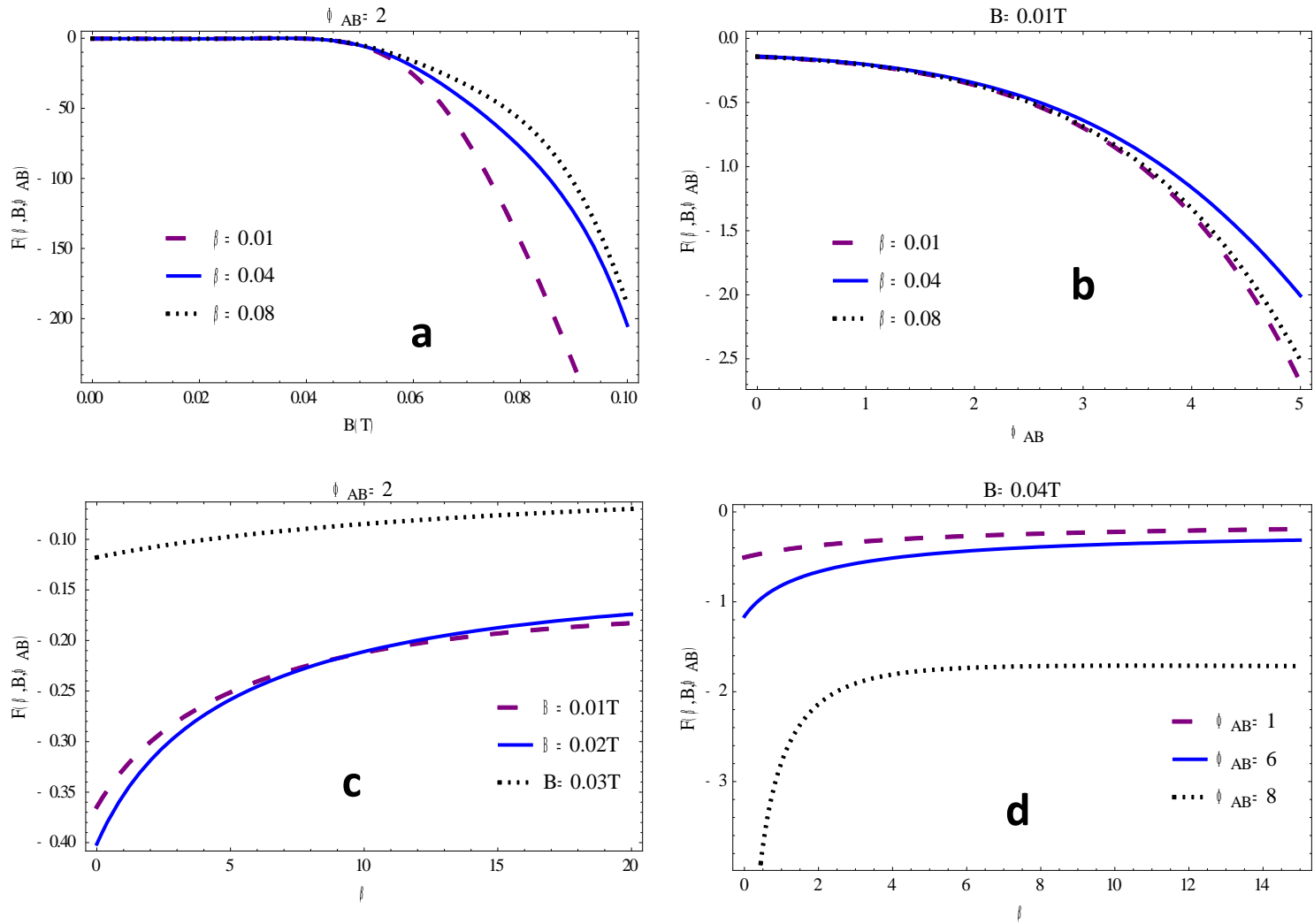

**Figure 7**; **(a)** Free energy, $F\left(\vec{B}, \Phi_{AB}, \beta\right)$ against $\vec{B}$-field varying $\beta$. **(b)** Free energy, $F\left(\vec{B}, \Phi_{AB}, \beta\right)$ against $\Phi_{AB}$ varying $\beta$. **(c)** Free energy, $F\left(\vec{B}, \Phi_{AB}, \beta\right)$ against $\beta$ varying $\vec{B}$-field. **(d)** Free energy, $F\left(\vec{B}, \Phi_{AB}, \beta\right)$ against $\beta$ varying $\Phi_{AB}$.

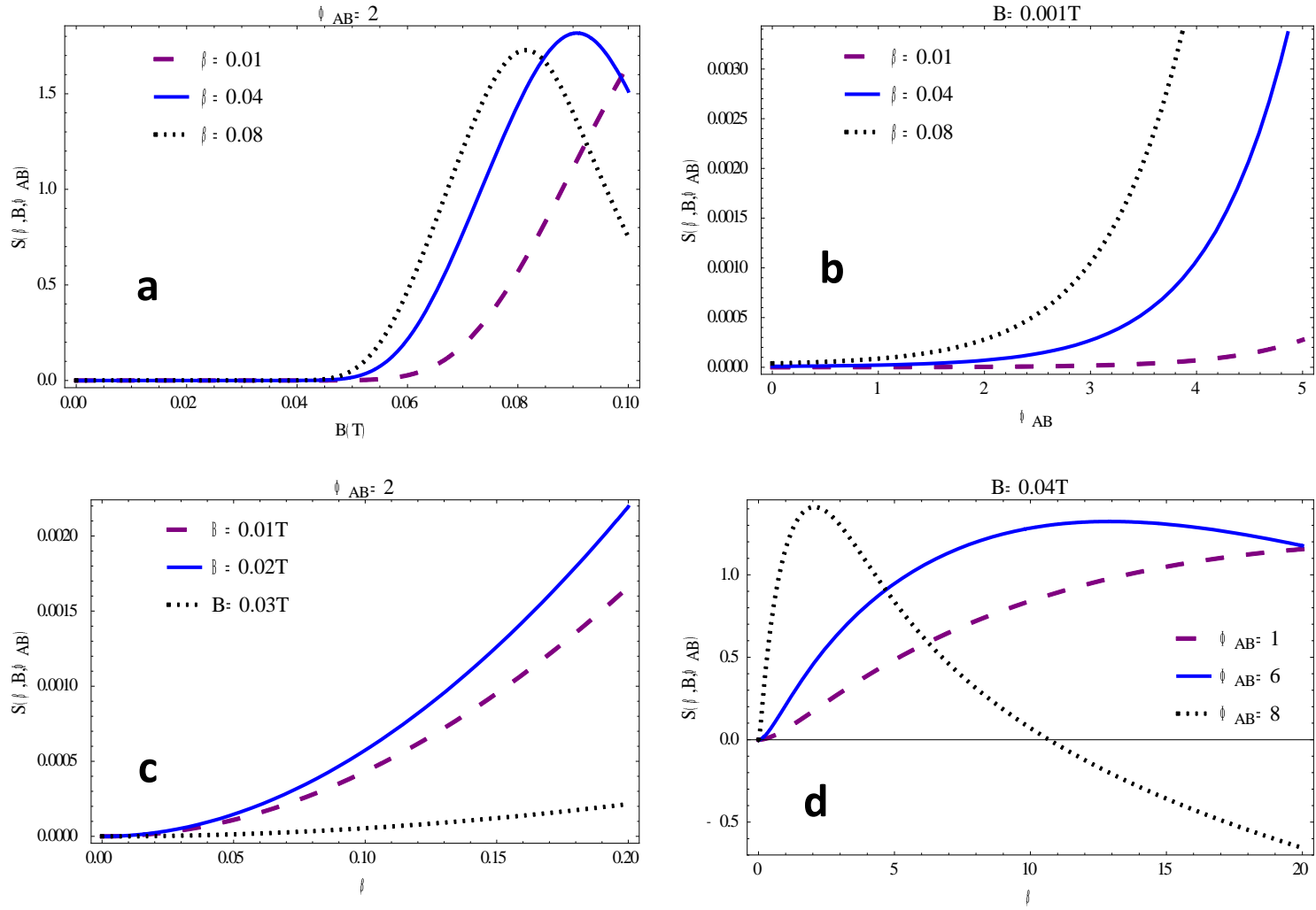

**Figure 8**; **(a)** Entropy, $S(\vec{B},\Phi_{AB},\beta)$ against $\vec{B}$-field varying $\beta$. **(b)** Entropy, $S(\vec{B},\Phi_{AB},\beta)$ against $\Phi_{AB}$ varying $\beta$. **(c)** Entropy, $S(\vec{B},\Phi_{AB},\beta)$ against $\beta$ varying $\vec{B}$-field. **(d)** Entropy, $S(\vec{B},\Phi_{AB},\beta)$ against $\beta$ varying $\Phi_{AB}$.

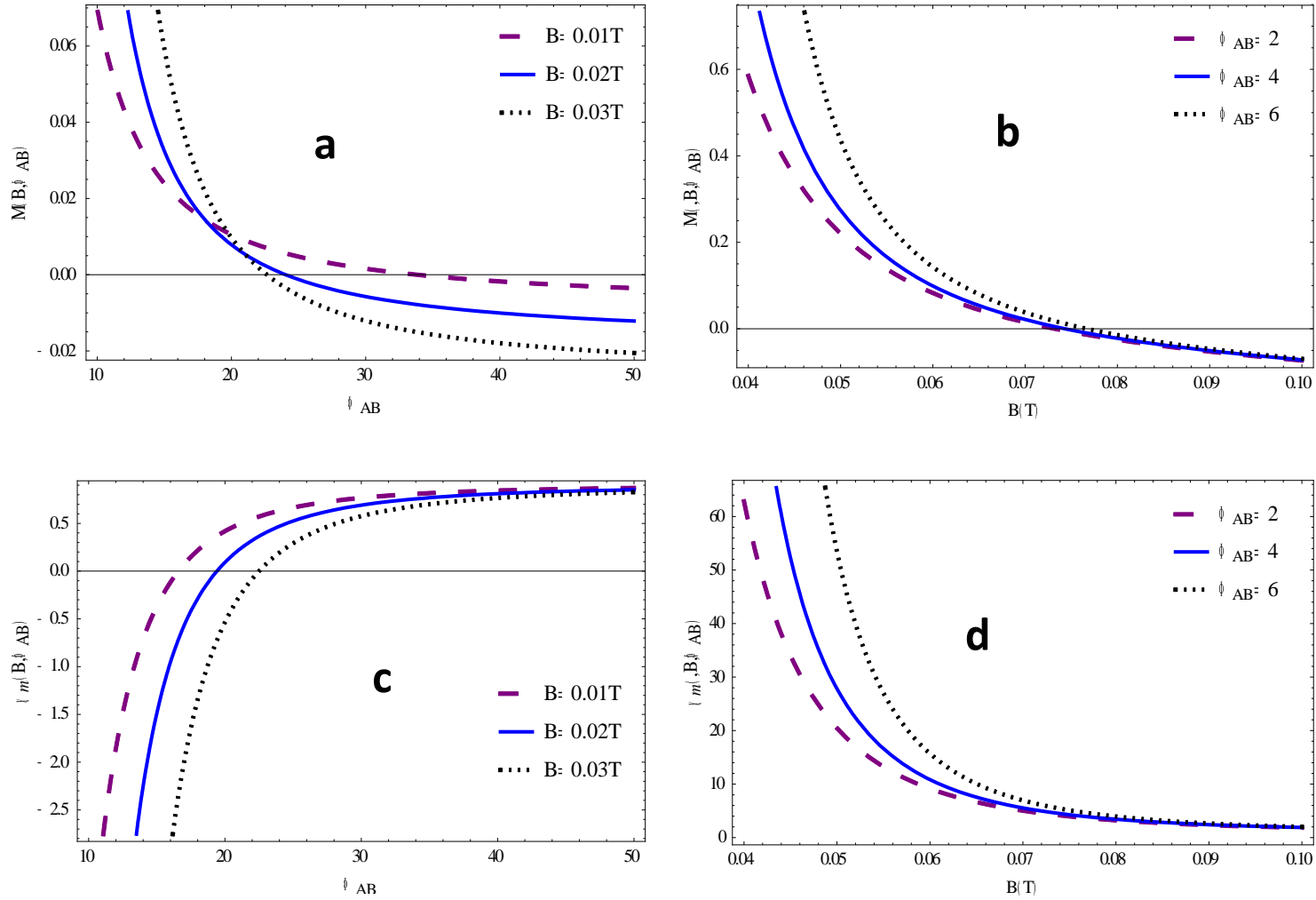

**Figure 9**; **(a)** Plot of Magnetization against AB flux field for different values of $\vec{B}$-field at zero temperature. **(b)** Plot of Magnetization against $\vec{B}$-field for different values of AB flux field at zero temperature. **(c)** Plot of Magnetic Susceptibility against AB flux field for different values of $\vec{B}$-field at zero temperature. **(d)** Plot of Magnetic Susceptibility against $\vec{B}$-field for different values of AB flux field at zero temperature.